\documentclass{PoS}

\title{Discovery of a nearby long, soft gamma-ray burst with an associated supernova}

\ShortTitle{Discovery of a gamma-ray burst with an associated supernova}

\author{\speaker{Rhaana L.C. Starling}\\
        University of Leicester\\
        E-mail: \email{rlcs1@star.le.ac.uk}}

\author{K. Wiersema\\
        University of Leicester\\
        E-mail: \email{kw113@star.le.ac.uk}}
\author{A.J. Levan\\
        University of Warwick\\
        E-mail: \email{a.j.levan@warwick.ac.uk}}

\abstract{We report the discovery of the nearby long, soft GRB 100316D, and the subsequent unveiling of its host
galaxy and associated supernova. We study the extremely unusual prompt emission with time-resolved gamma-ray to X-ray spectroscopy and find that a thermal component in addition to the synchrotron spectrum is required. The host galaxy is a bright, blue galaxy with a highly disturbed morphology. From optical photometry and spectroscopy we provide an accurate astrometry and redshift, and derive the key host properties of star formation rate and stellar age.
We compare our findings for this GRB-SN with the well known previous case of GRB\,060218. GRB 100316D is an important addition to the current sparse sample of spectroscopically confirmed GRB-SNe, from which a better understanding of long GRB progenitors and the GRB-SN connection can be gleaned.}

\FullConference{8th INTEGRAL Workshop ``The Restless Gamma-ray Universe''- Integral2010,\\
		September 27-30, 2010\\
		Dublin Ireland}

\begin{document}
\section{Introduction}
\label{sec:intro}
Gamma-Ray Bursts (GRBs) are proving a versatile tool with which to probe a number of the big open questions of astronomy today. GRBs are immensely luminous, and hence can be seen out to great distances: the current record-holder lies at $z=8.2$ [1]. GRBs pinpoint the most distant galaxies which host them, that would otherwise have remained unknown to us, and are a unique probe of conditions in the early Universe. Long-duration GRBs are thought to signal the deaths of massive stars, onset of core-collapse supernova events and the births of black holes. The GRB emission arises from a highly relativistic jet - extreme physical conditions that are probed through multiwavelength observational campaigns. This origin in massive stars, the latest evidence for which comes from GRB\,100316D, makes GRBs probes of the star formation history at all epochs.

The connection between long-duration GRBs and Type Ic core-collapse supernovae (SNe) has long been established, beginning with the association of GRB\,980425 with SN\,1998bw [2,3]. Spectroscopically confirmed examples of a handful of nearby GRB-SN associations have so far been found out to $z = 0.17$. In addition, a dozen or more GRBs out to $z\sim1$ have associated supernovae that are identified via a characteristic `bump' in the photometric data [4]. However, the majority of GRBs lie at higher redshifts ($\langle z \rangle = 2.2$ [5]) where such signatures would be impossible to detect. The GRBs with SNe are therefore rare, but provide a crucial insight into the GRB--SN connection and the progenitors of long bursts.

The GRBs with associated SNe are generally underluminous and subenergetic in comparison to a typical long GRB, and have prompt emission spectra which peak at lower energies [6]. They are suggested to have less relativistic outflows, or to be viewed more off-axis. It has been proposed that the observed nearby ($z<0.1$), underluminous/subenergetic GRBs may be drawn from a different population to the cosmological GRB sample [7,8].

In 2006 the {\it Swift} satellite detected the landmark event GRB\,060218. It was observed to have an unusually long-duration, low-luminosity and soft spectral peak, via comprehensive multiwavelength coverage of the prompt emission. The prompt spectrum was found to comprise both the non-thermal synchrotron emission ascribed to most GRBs and thought to originate in the collision of fast-moving shells within the GRB jet [9], and a thermal component. The presence of a thermal component led to the suggestion that we were observing the shock breakout of the supernova for the first time [10,11]. However, the non-thermal emission did not differ greatly from that of the X-ray flash class of soft GRBs and an outflow speed close to the speed of light could be inferred [12], alternatively suggesting this to be an extension of the typical GRB population and not requiring significantly slower ejecta or any special (off-axis) geometry [13]. [12] speculate that  very long-duration, low-luminosity events like this one may point to a different central engine compared with more typical GRBs: a neutron star rather than a black hole (see also [14,15]). 

We report here the discovery of a new GRB-SN, GRB\,100316D associated with SN\,2010bh [16,17]. Similarly to GRB\,060218, this is an unusually long-duration, soft-spectrum GRB positioned on a nearby host galaxy. We report precision astrometry for the source and the redshift of the underlying host galaxy. We examine the early GRB emission observed with the {\it Swift} satellite, and the broad characteristics of the host galaxy as observed with the Gemini South telescope, the Very Large Telescope (VLT) and the Hubble Space Telescope (HST). Studying this GRB-SN in detail, we seek to understand the origins of long-duration GRBs.

\section{Observations}
On 2010 March 16 at T$_{\rm 0}$ = 12:44:50 UT, the {\it Swift} Burst Alert Telescope (BAT) triggered on and
slewed immediately to GRB\,100316D. The X-Ray Telescope (XRT) and UV-Optical Telescope (UVOT) on-board began data-taking $\sim$140~s later. XRT found a bright source, decaying slowly with $\alpha = 0.13\pm0.03$ out to T$_{\rm 0}$+737 s. Only the host galaxy was detected with UVOT. 
Ground-based optical observations of GRB\,100316D were obtained with Gemini South, starting on March 16
23:53 UT and spanning several nights, in which we discovered the brightening supernova shown in Figure 1 [16]. We also acquired deep imaging with FORS2 and spectroscopy with the 3-arm
echelle spectrograph X-shooter, both mounted on the VLT.
Finally, we triggered HST, and report the first epoch obtained on March 25 also shown in Figure 1.
For full details of the observations and data analysis we refer to [18].
\begin{figure*}
\begin{center}
\includegraphics[width=5cm, angle=0]{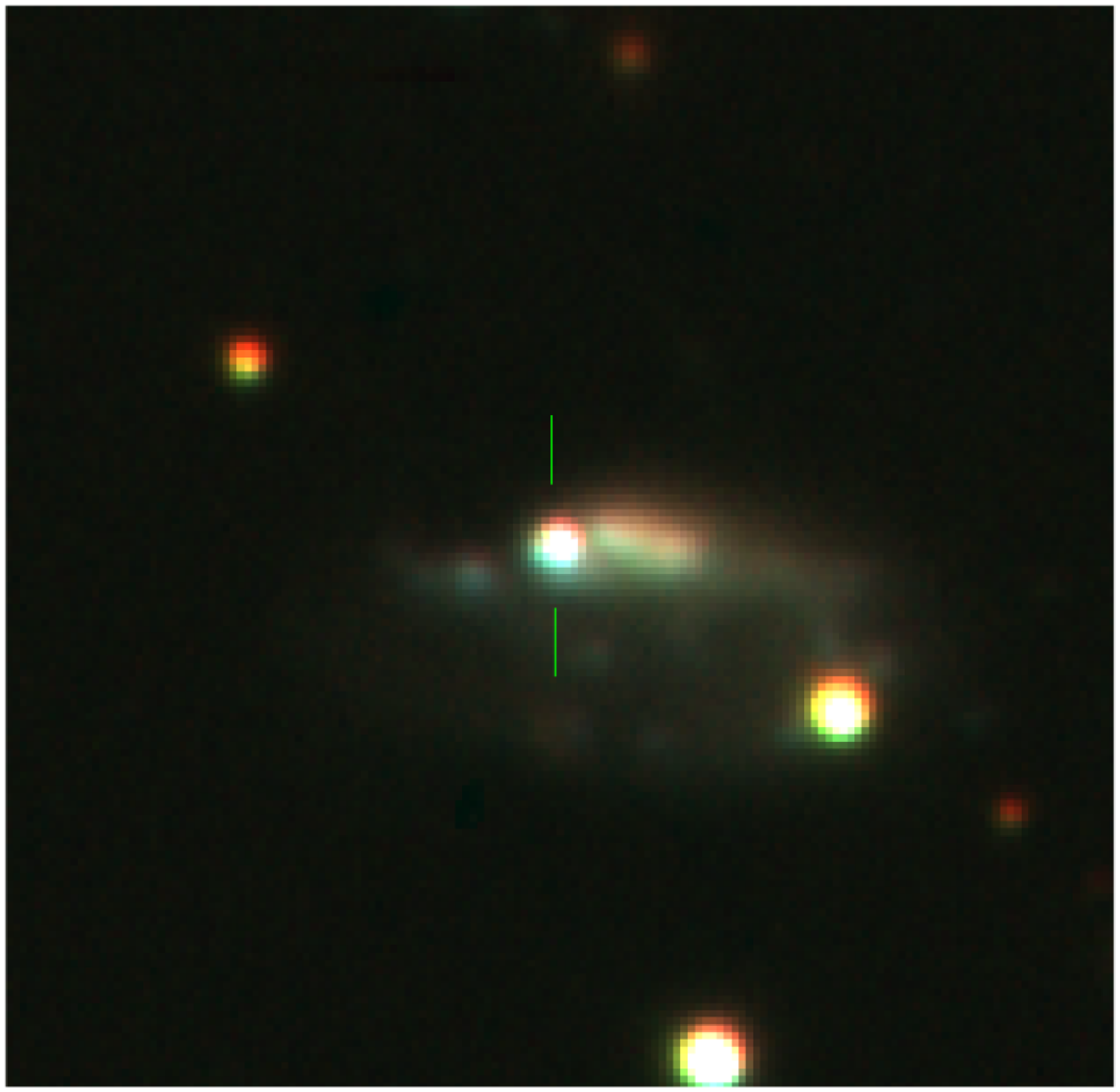}
\hspace{0.2mm}
\includegraphics[width=9cm, angle=0]{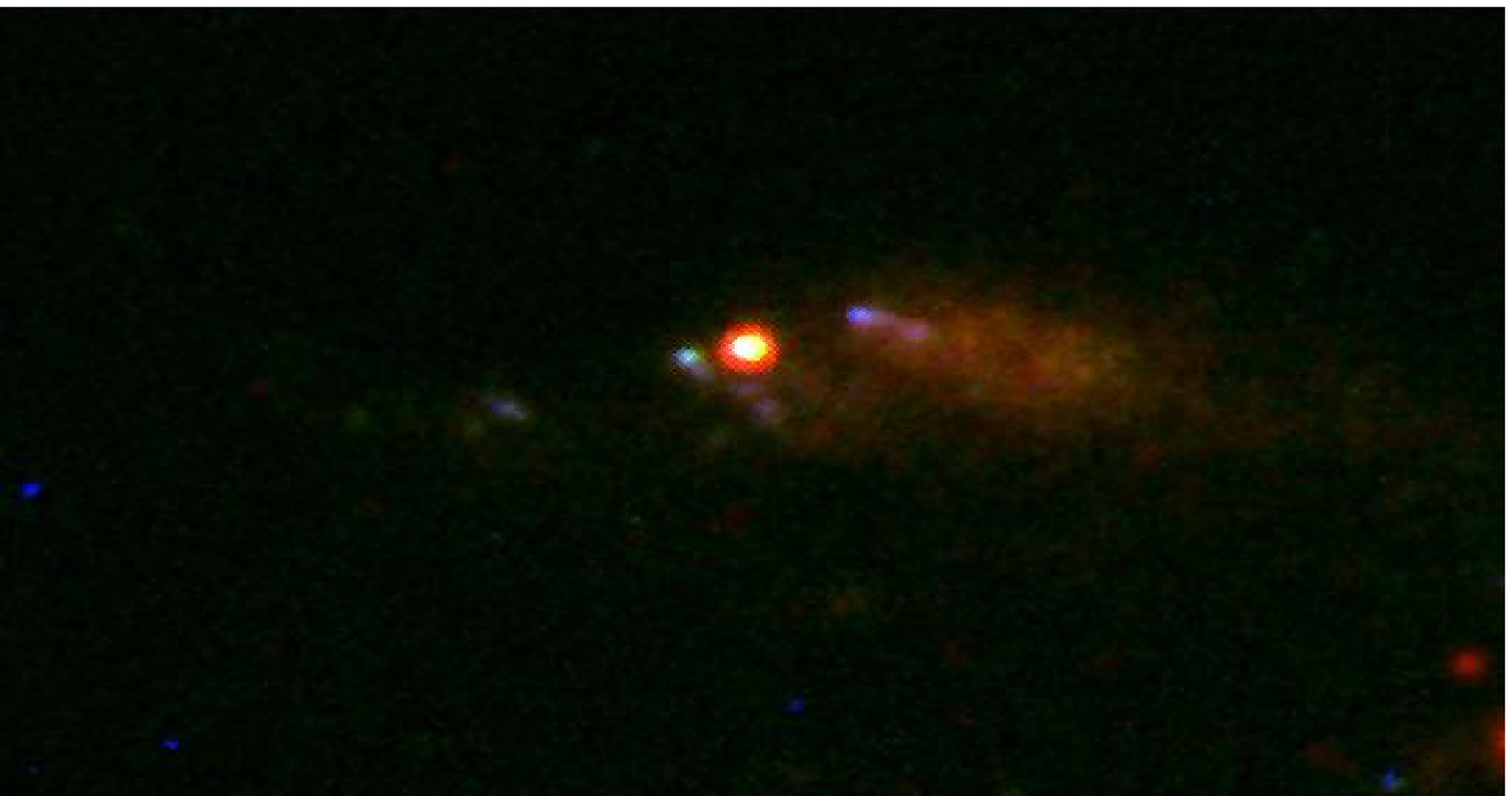}
\caption{Left: Gemini-South GMOS 3-colour image of the host galaxy of GRB\,100316D and its associated supernova SN\,2010bh (image $\sim$23$''$ across). Right: HST WFC3/UVIS image of the host galaxy. The disturbed morphology can be clearly seen.}
\label{subtraction}
\end{center}
\end{figure*}
\begin{figure}
\begin{center}
\includegraphics[width=5cm, angle=-90]{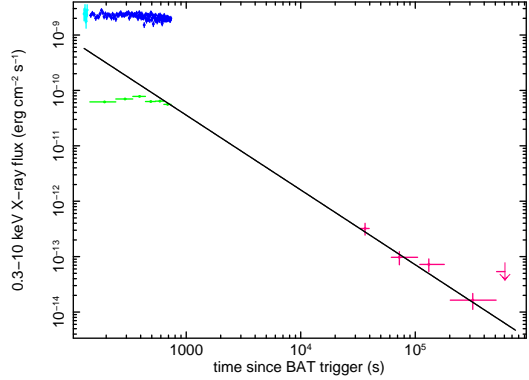}
\caption{Flux light curve for the {\it Swift} XRT X-ray data (cyan: Windowed Timing (WT) settling mode; blue: WT mode; pink: Photon Counting mode) with best fitting $\alpha= 1.35\pm 0.30$ power law model to the late-time data overlaid (solid black line). This decay clearly cannot be extrapolated back to the slowly decaying prompt emission. The green points indicate the contribution from a blackbody component.}
\label{XRTlc}
\end{center}
\end{figure}

\section{Results}
From our HST imaging, we determine a best position for the GRB-SN of RA, Dec (J2000) = 07h 10m 30.530s, $-$56d 15' 19.78$''$ ($\pm$0.05$''$).
We detect a large number of nebular emission lines, commonly found in
star forming regions, in the VLT X-shooter spectrum of an HII region of the host galaxy. From the 12 brightest lines we measure the redshift $z = 0.0591 \pm 0.0001$.

The prompt emission duration of this GRB is one of the longest ever measured (Figure 2). GRB\,100316D was detected with BAT from $\sim$T$_{\rm 0}$$-$500~s to at least $\sim$T$_{\rm 0}$+800~s, giving a lower limit on the duration of $\sim$1300~s. The isotropic equivalent energy that can be derived from the 15--150~keV spectrum is $E_{\rm iso} \ge$~(3.9$\pm$0.3)~$\times$10$^{49}$ erg. Extrapolating to the 1--150 keV band for this soft burst gives $E_{\rm iso} \ge$~(5.9$\pm$0.5)~$\times$10$^{49}$ erg. The BAT and XRT simultaneous coverage of 0.3--150 keV spans 603~s. We performed time-resolved spectroscopy during this overlap for the BAT and XRT data both seperately and jointly. The spectra are modelled with an exponentially cut-off power law typical of GRB prompt emission. We then included a thermal blackbody component, which significantly improves the fit in both cases (shown to be $>4\sigma$ significant in 10000 trials with a Monte Carlo analysis [18]). The blackbody component contributes $\sim$3\% of the total observed 0.3--10 keV X-ray flux (Figure 2), and has a luminosity of (3--4)$\times$10$^{46}$ erg s$^{-1}$ corresponding to an emitting radius of 8$\times$10$^{11}$~cm. The peak energy is at $\sim$30 keV in the first time interval, decreasing thereafter. We show the evolution of the power law photon index, spectral peak energy and blackbody temperature in Figure 3: remarkably similar to the evolution of the prompt emission spectrum of GRB\,060218 shown for comparison [6]. A full table of spectral fitting results is given in [18].

%\begin{figure}
%\begin{center}
%\includegraphics[width=6cm, angle=-90]{../plots/blackbodycontribution.ps}
%\caption{Unfolded XRT WT time-averaged spectrum, $EF_E$ vs. $E$, with an absorb%ed power law plus blackbody model, showing the contributions of the two compone%nts.}
%\label{XRTbbody}
%\end{center}
%\end{figure}
\begin{figure}
\begin{center}
\includegraphics[width=6cm, angle=0]{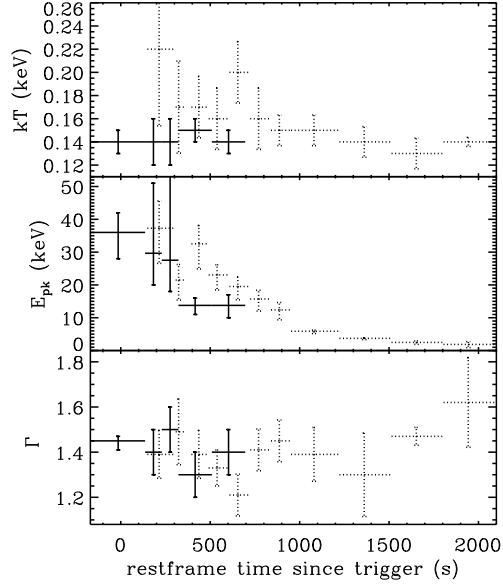}
%\hspace{0.2mm}
%\includegraphics[width=5cm, angle=-90]{../plots/blackbodycontribution.ps}
\caption{Evolution of power law photon index $\Gamma$, peak energy $E_{\rm pk}$ and blackbody temperature $kT$ in the blackbody plus exponentially cut-off power law model fitted to the BAT-XRT spectra for GRB\,100316D (solid error bars), compared with the same model fitted to BAT-XRT data of GRB\,060218 (taken from [6], dotted error bars) in the source rest-frames. Errors are 90\% confidence.}
\label{Epkevolcf060218}
\end{center}
\end{figure}

\subsection{The host galaxy}
The properties of GRB host galaxies can provide additional clues to the progenitor evolution, through determination of properties such as the age, metallicity and alpha element enrichment of the
stellar population. Low redshift GRB-SNe such as this one are of particular interest, as they allow spatially resolved spectroscopy to be obtained [19]. The derived properties can then be
meaningfully compared to those of `normal' core-collapse supernova hosts [20,21] or other star forming galaxy populations. 

Using the spectrum of an HII region in the host galaxy of GRB\,100316D, uncontaminated by the GRB and supernova, we compute some diagnostics of the galaxy.
We detect Balmer absorption lines underlying the nebular
emission, common in GRB host galaxy spectra. At sufficient resolution this can provide a useful age tracer (Wiersema et al. in preparation). We fit the H$\delta$ absorption component, and find an approximate age for
the dominant stellar population of $\sim30$ Myr, assuming continuous star formation. The detection of the bright H$\alpha$ line allows us to estimate, from its luminosity, the star formation rate. Using the de-reddened ($E(B-V) = 0.178$ mag, [18]) flux from the X-shooter spectrum we find $SFR_{{\rm H}\alpha} = 0.17$ M$_{\odot}$ yr$^{-1}$ (formally a lower limit). More details can be found in [18], and a full analysis of the host galaxy will appear in Flores et al. (in preparation).

\subsection{Comparison with GRB\,060218}
Both GRB\,100316D and its predecessor GRB\,060218 are nearby ($z=0.059, 0.033$), long-duration ($T_{90} \ge$1300~s,~2100~s), initially relatively constant in X-ray flux (Figure 1), spectrally soft (very few or no counts above 100 keV; low $E_{\rm pk}$, Figure 3), subenergetic (both have $E_{\rm iso}\sim 4\times 10^{49}$ erg) GRBs with a spectroscopically confirmed associated SN. These two events show similar prompt emission properties and stand out among the GRB-SNe subsample for their unusual X-ray evolution and the need for a thermal X-ray emission component (see e.g. [6],[10]). However, their host galaxies are rather different in morphology and metallicity [22,18], with the host of GRB\,100316D more closely resembling the host of GRB\,980425 (see [18],[2],[3]).

\section{Conclusions}
GRB\,100316D is an atypical gamma-ray burst both in its temporal and spectral behaviour. The very soft spectral peak and extended and slowly decaying flux emission are highly unusual among the prompt emission of GRBs [23]. The discovery of a thermal component in the {\it Swift} XRT X-ray spectra, seen only once before in a GRB-SN [10], may be the signature of shock breakout of the supernova itself; this is unlikely to be confirmed without further examples of this phenomenon.
The dominant component of the high energy emission in GRB\,100316D remains the synchrotron-like non-thermal spectrum common to all types of GRB (with and without SNe) thought to originate in internal shocks in a relativistic jet. The long duration of the early X-ray emission is curious, and exceedingly rare, perhaps suggesting a greater reservoir of material is available to feed the central engine and prolong its activity.

\section{Acknowledgements}
The authors would like to thank the wider collaboration who contributed to this effort. This work made use of data supplied by the UK {\it Swift} Science Data Centre at the University of Leicester. Based on observations made with ESO Telescopes at the La Silla or Paranal Observatories under
programme IDs 084.D-0939, 083.A-0644 and 084.A-0260(B). Based on observations made with the NASA/ESA Hubble Space Telescope, obtained at the Space Telescope Science Institute, which is operated by the Association of Universities for
Research in Astronomy, Inc., under NASA contract NAS 5-26555: these observations are
associated with programme 11709. We thank STScI staff for their efforts in implementing the HST ToO observation,
particularly Alison Vick. RLCS and KW acknowledge financial support from STFC. Financial support of the British Council and Platform Beta Techniek through the Partnership Programme in Science (PPS WS 005) is also gratefully acknowledged (KW).

\end{document}